\documentclass{ws-procs9x6}

\begin{document}

\title{Results on strangeness production from HADES}

\author{A. Schmah$^*$ and L. Fabbietti for the HADES collaboration}

\address{Excellence Cluster Universe, Technical University Munich,\\
Garching, 85748, Munich\\
$^*$E-mail: a.schmah@gsi.de\\
www.universe-cluster.de}

\begin{abstract}
Resent results concerning the production of $\mathrm{K^+}$, $\mathrm{K^-}$, $\phi$, $\mathrm{K^{0}_{S}}$ and  $\Lambda$ particles
in the reaction system Ar+KCl at 1.76 AGeV measured by the HADES Collaboration are presented. Transverse momentum distributions have
been measured in a large fraction of the phase space.
For the first time, at SIS energies, $\mathrm{K^+}$, $\mathrm{K^-}$ and  $\phi$ mesons have been measured independently and the slope
parameter for  $\phi$ mesons could be extracted. A large statistics has been collected for the produced $ K^{0}_{S}$ and  $\Lambda$
in the phase space region around mid-rapity. The high quality of these data render them very suitable for comparisons with theoretical models.

\end{abstract}


\bodymatter

\section{Introduction}\label{sec:0}
The study of hadron properties in baryonic matter has been pursued via several experiments in the last two decades.
The systematic investigation of $\mathrm{K^\pm}$ sub-threshold production yields, phase space distributions
and flow observables in relativistic heavy-ion collisions at various beam energies (1-2 AGeV) and for
various system sizes (from C+C to Au+Au) and centralities has attracted much attention, related to distinct
in-medium properties of $\mathrm{K^+}$ and $\mathrm{K^-}$ mesons
\cite{Kaplan86,Brown94,Hartnack94,Ko96,Li97,Cassing97}.
Corresponding experiments have been the focus of previous studies by the KaoS
\cite{Sturm01,Foerster03,Senger04,Uhlig05,Foerster07}
and FOPI \cite{Ritman95,Best97,KWisnia00,Crochet00,Mangiaro03}
collaborations at GSI. From the analysis of the out-of-plane \cite{Shin98,Li96} and sideward flow \cite{Crochet00,Brat97} the $\mathrm{K^+}$-nucleon potential is expected to be repulsive and the $\mathrm{K^+}$ measured production yields are in agreement with the scenario of a soft nuclear equation of state ($\mathrm{K_N}\approx 200~MeV$) \cite{Fuchs01,Hartnack02}.\\
Recently published KaoS data \cite{Foerster07} gave an important contribution
towards understanding the production mechanism of $\mathrm{K^{\pm}}$ mesons.
Indeed, the combined analysis of $\mathrm{K^+}$ and $\mathrm{K^-}$ yielded the interpretation
that a substantial part of the observed
$\mathrm{K^-}$ mesons is due to a strangeness exchange mechanism \cite{Hartnack03}. \\
On the other hand, attention has been drawn to the study of $\mathrm{K^{0}_{S}}$ and $\Lambda$ production in heavy collision for the mentioned energy range.  The study of the neutral strange particles could deliver information on the in-medium potential without the contribution of the Coulomb interaction \cite{Benab08}. 
The spectrometer HADES \cite{hadesSpectro}, primarily designed to measure
di-electrons \cite{HADES-PRL07}, has been recently employed
for the identification of strange mesons as well, showing high purity and efficiency
for particle identification and excellent reconstruction capability
of secondary decay vertices \cite{PhD_Schmah}.
For the first time, a combined and exclusive identification of
$\mathrm{K^+}$, $\mathrm{K^-}$ and $\phi$ mesons was carried out at sub-threshold beam energy for the reaction Ar +KCl at 1.76 AGeV \cite{Aga09}.
(``Sub-threshold'' refers to free nucleon-nucleon collisions and is related here
to $\mathrm{K^-}$ and $\phi$ channels.)
Our measurement allows firm conclusions on the fraction of $\mathrm{K^-}$ mesons stemming from $\phi$ decay \cite{Aga09}. \\
At the same time the reconstruction of the $\mathrm{K^{0}_{S}}$ and $\Lambda$ particles was carried out. The high statistics collected for these particles in the phase space region of mid-rapidity render these data particularly interesting for further comparisons with transport models.

\section{Spectrometer setup}\label{sec:1}
The experiment was performed with the
{\bf H}igh {\bf A}cceptance {\bf D}i-{\bf E}lectron {\bf S}pectrometer HADES
at the heavy-ion synchrotron SIS (SchwerIonenSynchrotron)
at GSI (Gesellschaft f\"ur Schwerionenforschung)
in Darmstadt, Germany. A detailed description of the spectrometer is presented in
\cite{hadesSpectro}. HADES consists of a 6-coil toroidal magnet centered on the beam axis and
six identical detection sections located between the coils and covering
polar angles between $15^{\circ}$ and $85^{\circ}$.
The sectors were composed of a gaseous Ring-Imaging Cherenkov (RICH) detector, four planes of
Multi-wire Drift Chambers (MDCs) for track reconstruction, and two
Time-of-Flight walls (TOF/TOFino) supplemented at forward polar angles
with Pre-Shower chambers. \\
An argon beam of $\sim 10^6$ particles/s was incident on a four-fold segmented
KCl target with a total thickness corresponding to $3.3 \%$
interaction length. A fast diamond start detector located upstream
of the target was used to determine the interaction time.
The data readout was started by a first-level trigger (LVL1) decision,
requiring a charged-particle multiplicity, $MUL \ge 18$, in the TOF/TOFino detectors,
accepting approximately 35\,\% of the nuclear reaction cross section.

\section{Particle identification}

\begin{figure}
\centering
\resizebox{0.8\textwidth}{!}{
\includegraphics{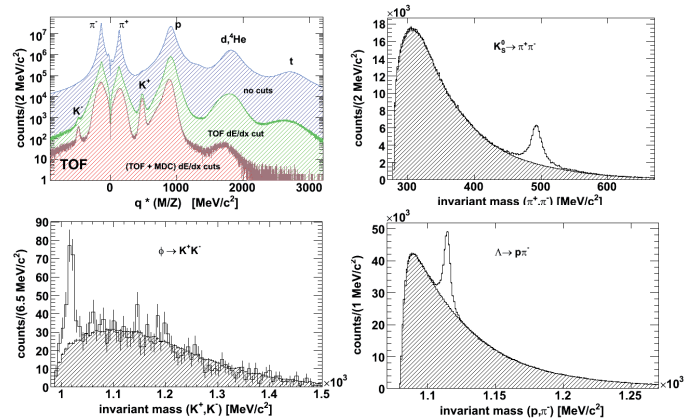}
}
\caption[]{Top left: Polarity times mass over charge number distributions for particle tracks measured in the TOF detector for different cut conditions.
The spectrum on the top (blue curve) shows the original distribution,
whereas the lower two spectra show the effects of the graphical cuts
in the TOF $dE/dx$ (green histogram) and MDC $dE/dx$ (blue histogram) distributions. Invariant mass distributions of $\mathrm{K^+}$ and $\mathrm{K^-}$ pairs (bottom left), $\mathrm{\pi^+}$ and $\mathrm{\pi^-}$ pairs (top right) and $\mathrm{p}$ and $\mathrm{\pi^-}$ pairs (bottom right). The grey shaded areas show the combinatorial background calculated with the mixed event technique.
 \label{Kaon_mass}}
\end{figure}

Hadrons have been identified using the time-of-flight walls TOF and TOFino(+Pre-Shower) and the MDCs.
The particles can be distinguished by their velocity and by their energy loss ($dE/dx$) 
in one of these detectors as a function of their momentum. For the hadron identification graphical cuts were applied around these distributions \cite{PhD_Schmah}.  \\
For the $\mathrm{K^{\pm}}$ identification only the TOF data are considered,
because of rather large background in the TOFino due to its limited granularity and time
resolution. In order to drastically reduce the background of protons and pions, graphical cuts according to the calculation of the TOF $dE/dx$ and the MDC
$dE/dx$ distributions have been applied. Figure \ref{Kaon_mass} (top left) shows the polarity times mass over charge number distributions for particle tracks measured in the TOF detector without any {dE/dx} cut and when the TOF and TOF+MDC $dE/dx$ cuts for  $\mathrm{K^+}$ and $\mathrm{K^-}$ are applied. One can see how the kaon peaks stick out of the background. \\
The $\mathrm{K^+}$ and $\mathrm{K^-}$ yields are extracted applying a Gaussian fit of the signal after the background subtraction and the integration
over the whole phase space \cite{Dip_Lorenz}. In order to reconstruct $\phi$ mesons, $\mathrm{K^+}$ and $\mathrm{K^-}$, are combined to pairs.\\
The resulting invariant mass distributions for  $\mathrm{K^+}-\mathrm{K^-}$, $\pi^+-\pi^-$ and $p-\pi^-$ pairs are shown in
figure~\ref{Kaon_mass} where $\phi$, $\Lambda$ and $K_{S}^{0}$ peaks are clearly visible. The combinatorial background was described using the mixed event technique.

\section{Transverse mass spectra and inverse slopes}
\label{trmass}
The geometrical acceptance for the particles has been determined
as a function of the emission rapidity and transverse mass.
A full-scale simulation of the HADES spectrometer was performed with the
GEANT3 package \cite{geant}. The reconstruction efficiency for the particles has been determined using simulations, exploiting the embedded tracks method \cite{Aga09}.\\
\begin{figure}
\centering
\resizebox{0.7\textwidth}{!}{
\includegraphics{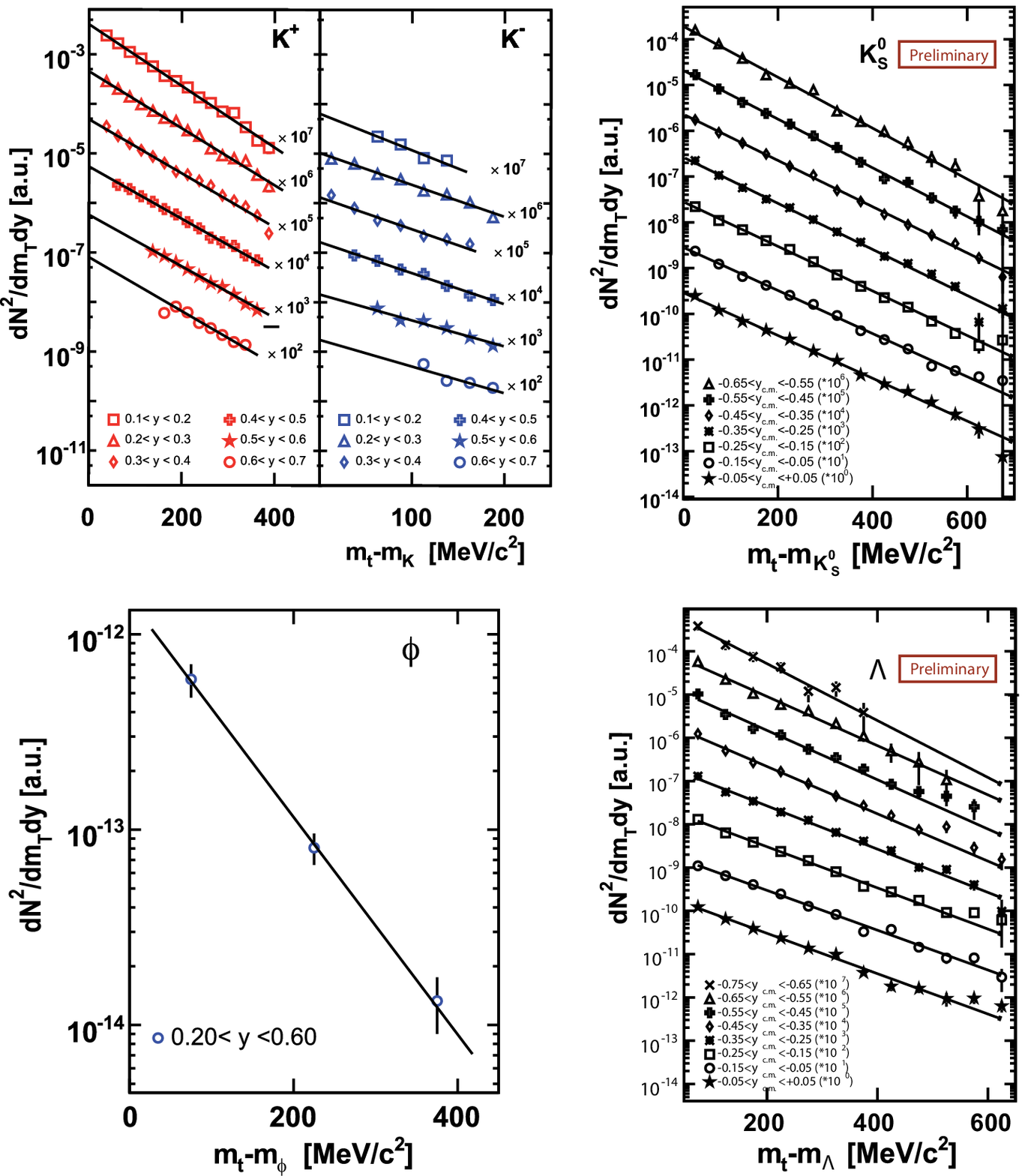}
}
\caption[]{Transverse mass spectra of reconstructed $\mathrm{K^+}$, $\mathrm{K^-}$, $\phi$, $\Lambda$ and $K_{S}^{0}$.
The spectra are plotted for several laboratory rapidity bins.
The spectra are scaled by the factors as indicated in the plots.
\label{mtKaons}}
\end{figure}
The efficiency corrected transverse mass spectra for $\mathrm{K^+}$, $\mathrm{K^-}$, $\Lambda$, $\mathrm{K^0_S}$ and $\phi$ for different laboratory rapidity bins
are exhibited in figure~\ref{mtKaons}. If a Boltzmann fit according to
$\frac{1}{m_{t}^{2}} \frac{d^2N}{dm_{t}dy} = C(y) \,
\exp \left( -\frac{(m_t-m_0)c^2}{T_B(y)}  \right) $
is applied, a good description of the experimental data is achieved. \\
The so obtained inverse slope parameters $T_B(y)$ for each rapidity bin $y$ are used to extract an effective inverse slope parameter of all particle species. This is done by assuming a thermal emission source and fitting the  $T_B(y)$ distribution \cite{Aga09} according to
$T_B(y) = \frac{T_{\mathrm{eff}} } {\cosh(y)}$.
The obtained parameter $T_{\mathrm{eff}}$ represents the inverse slope at cm rapidity
and can be considered as an effective temperature of the kinetic freeze-out of the
respective particle. The  $T_{\mathrm{eff}}$  values obtained for the particles are shown in table \ref{TempMulti}. \\
One can see that $T_{\mathrm{eff}}$ is lower for $\mathrm{K^-}$ than for $\mathrm{K^+}$.
This can be understood as different freeze-out conditions for the two meson species
\cite{Foerster03}. \\ 
If we consider the $T_{\mathrm{eff}}$ obtained for the $\phi$ meson, one can see that this value is much higher than the one obtained for $\mathrm{K^-}$ and slightly smaller than the $\mathrm{K^+}$ value. Since the event statistics is not sufficient to tag the fraction of kaons which are produced in the $\phi$ decay, it is difficult to draw conclusions. One might interpret this difference in the temperature again as a signature of the different freeze-out conditions between $\phi$ and $\mathrm{K^-}$. If a considerable fraction of the  $\mathrm{K^-}$ is generated in secondary processes, the effective temperature might be lower. Following these considerations it would be very interesting to enlarge the event statistics and select exclusively on the $\mathrm{K^-}$ produced in the $\phi$ decay. \\
One can see that the low-momentum component of the  $\mathrm{K^0_S}$ spectrum has been measured with high statistics in the whole phase space, even in the mid-rapidity region. Since the mesons with lower momenta are supposed to experience more strongly the effect of the interaction potential with the nuclear medium, these data are very well suited to study in-medium effects.\\

\section{ Summary}
The high capability of the HADES spectrometer to be employed in the analysis of strange baryon and mesons has been demonstrated.
The good quality of the extracted signals for $\mathrm{K^+}$ and $\mathrm{K^-}$
has been shown via the mass, transverse mass and rapidity distributions.
A total of $168 \pm 18$ counts has been collected for
the $\phi$ meson in the reaction Ar + KCl at kinetic beam energy of 1.756 AGeV.\\
High event statistics and high quality data for the $\mathrm{K^0_S}$ and $\Lambda$ signals have been reconstructed over a large fraction of the phase space, including also the mid-rapidity region.
These data are well suited for the comparison with model calculations. In particular, the low momentum component of the $\mathrm{K^0_S}$ spectra should be sensible to in-medium properties of the interaction potential.
\begin{table}[hbt]
\begin{center}
\begin{tabular}{|c|c|}
    \hline
    Particle & $T_{Eff}~ [MeV]$ \\
    \hline
    $\Lambda+\Sigma^0$  & $95.8  \pm 0.8  \pm 0.5$  \\
    \hline
    $\mathrm{K^+}$  & $89 \pm1 \pm2$\\
    \hline
    $\mathrm{K^{0}_{S}}$ & $92.0 \pm 0.5\pm4.1$\\
    \hline
    $\mathrm{K^-}$  & $69 \pm2 \pm4$\\
    \hline
    $\phi$   & $84 \pm8 $\\
    \hline
\end{tabular}
\caption{Effective inverse slope parameters.
    }
    \label{TempMulti}
\end{center}
\end{table}

\bibliographystyle{ws-procs9x6}

\end{document}